\documentclass[]{raa}            % referee version: for submission

\usepackage{graphicx,times}             %for PS/EPS graphics inclusion, new
\usepackage{natbib}
\usepackage{amssymb,amsmath}
\usepackage[pagebackref=true]{hyperref}
\hypersetup{colorlinks = true, linkcolor = green, anchorcolor = red, citecolor = blue, filecolor = red, urlcolor = red}

\usepackage{ulem}

\begin{document}

   \title{Photometric and Spectroscopic Study of Flares on Ross 15
%\footnotetext{$*$ Supported by the National Natural Science Foundation of China.}
}
%   \subtitle{I. Place Your Subtitle Here}

   \volnopage{Vol.0 (20xx) No.0, 000--000}      %%preserved for Editor. DOn't remove!
   \setcounter{page}{1}          %%starting page, preserved for Editor. DOn't remove!

   \author{Jian-Ying Bai
      \inst{1,2}
   \and Ali Esamdin
      \inst{1}
   \and Xing Gao
      \inst{1}
   \and Yan Yan
      \inst{3,4}
   \and Juan-Juan Ren
      \inst{3}
   }

   \institute{Xinjiang Astronomical Observatory, Chinese Academy of Sciences, Urumqi 830011, Xinjiang, People's Republic of China; {\it aliyi@xao.ac.cn}\\
        \and
             University of Chinese Academy of Sciences, Beijing 100049, People's Republic of China;\\
        \and
             National Astronomical Observatory of China, Chinese Academy of Sciences, Beijing, 100101, People's Republic of China;\\
        \and
             CAS Key Laboratory of Solar Activity, National Astronomical Observatories, Chinese Academy of Sciences, Beijing, 100101, People's Republic of China;\\
\vs\no
   {\small Received~~20xx month day; accepted~~20xx~~month day}}

\abstract{We conducted photometric and spectroscopic observations for Ross 15 in order to further study the flare properties of this less observed flare star. A total of 28 B-band flares are detected in 128 hours of photometric observations, leading to a total flare rate of 0.22$^{+0.04}_{-0.04}$ hour$^{-1}$, more accurate than that provided by previous work. We give the energy range of the B-band flare (10$^{29.5}$ - 10$^{31.5}$ erg) and the FFD for the star. Within the same energy range, the FFD are lower than that of GJ 1243 (M4) and YZ CMi (M4.5), roughly in the middle of those of three M5-type stars and higher than the average FFDs of spectral types $\geq$ M6. We performed, for the first time to Ross 15, simultaneous high-cadence spectroscopic and photometric observations, resulting in detection of the most energetic flare in our sample. The intensity enhancements of the continuum and Balmer lines with significant correlations between them are detected during the flare, which is same with that of the other deeply studied flare stars of the similar spectral type.
}

   \authorrunning{J.-Y. Bai et al.}            %author_head in even pages
   \titlerunning{Flare study with photometry and spectrometry }  % title_head in odd pages

   \maketitle
%% The author head (on even pages) and the title head (on odd pages) will be
%% automatically extracted from \author{} and \title{}. Whenever the title is too long,
%% you will be asked to supply a shorter one by inserting either \authorrunning{} or
%% \titlerunning{} before \maketitle. Anyway, you can specify your own heads.
%%
%%
%% Note: In the following text body of your manuscript, please note several differences from
%%       other major journals:
%% (1) \subsection{Please Capitalize the First Letter of Each Notional Word in Subsection Title}
%% (2) Please Capitalize the First Letter of Each Notional Word in all tables' captions

%
%________________________________________________ sections below
%
\section{Introduction}           %% first-level sections will be auto-capitalized
\label{sect:intro}

Flares are transient and violent phenomena occurring on the sun and stars where a large amount of energies are released over a wide range of wavelengths, from radio to X-rays \citep{2002A&A...394..653S,2006ApJ...647.1349O,2007ApJS..173..673W,2010ApJ...721..785O,2012ApJ...748...58D}. Stellar flares are considered to be triggered by magnetic reconnection in the corona \citep{2000IrAJ...27..117G,2005Ap.....48..279T,2010ARA&A..48..241B,2019ApJ...871L..26F}. Some last as short as 2 - 3 seconds or a few minutes, in which there exists an impulsive rise followed by a gradual exponential decay \citep{1989MmSAI..60..263G}. Therefore, using high-temporal resolution photometry and spectrometry will help to improve our understanding of their physical process \citep{2017ApJ...849...36Y,2018PASJ...70...62H,2019ApJ...876...58N}. Several studies have been published with photometric and simultaneous high-cadence spectroscopic observations \citep{2013A&A...560A..69L,2013ApJS..207...15K,2016ApJ...820...95K,2019ApJ...871L..26F}.

Ross 15 is a late-type dwarf with spectra type of dM4e \citep{1974ApJS...28....1J}. Recently, \cite{2018yCat..51550180M} reported for this star a Vmag of 12.155 $\pm$ 0.033 magnitude, a mass of 0.422 $\pm$ 0.004 M$_{\sun}$ , an effective temperature of 3253 $\pm$ 69 K and a radius of 0.390 $\pm$ 0.012 R$_{\sun}$. And, \cite{2018yCat..36160012G} gave its distance of 76.1335 $\pm$ 0.0541 mas (13.1348 $\pm$ 0.0093 pc).

Ross 15 was reported as a flare-star candidate in early research \citep{1955PASP...67...34R}. \cite{1977A&AS...30..113P} firstly studied the flare of Ross 15, reported three flares of it through B filter in 5.85 hours of observation, providing its flare rate of 0.51$^{+0.29}_{-0.29}$ hour$^{-1}$. No further study related to this star has been conducted so far.

To further investigate the properties of the flares on Ross 15, we conducted in this work a relatively long term of photometric observations and a period of simultaneous spectral and photometric observations to this star. This paper is organized as follows. In Section 2, the observations and the data reduction are outlined. In Section 3, we describe the method of data analysis for the photometry and the spectrometry. In Section 4, we present the results and give discussions in Section 5. In the final section, Section 6, we give the summary.

\begin{table}[ht]
\begin{center}

\caption[]{ \centering {Log of photometry} }\label{Tab:t1}
\begin{tabular}{cccccc}
  \hline\hline\noalign{\smallskip}
Date &  Filter      & Start & End & Duration & Flare \\
(UT) &  (Johnson)    & (UT) & (UT) & (Hour) & Number \\
  \hline\noalign{\smallskip}
2018 Oct 21 & B& 17:56:39 & 23:29:53 & 5.42& 2		\\
2018 Oct 26 & B& 14:17:08 & 23:29:32 & 9.2& 3		\\
2018 Nov 9 & B& 16:43:33 & 22:27:56 & 5.73& 1		\\
2018 Nov 10 & B& 16:00:21 & 23:39:59 & 7.65& 1		\\
2018 Nov 22 & B& 14:16:28 & 19:08:43 & 4.87& 0		\\
2018 Nov 26 & B& 14:44:12 & 23:49:39 & 9.08& 1		\\
2018 Nov 28 & B& 17:08:55 & 23:49:57 & 6.68& 1		\\
2018 Dec 11 & B& 14:40:13 & 19:57:58 & 5.28& 2		\\
2018 Dec 14 & B& 18:25:33 & 23:11:00 & 4.77& 0		\\
2018 Dec 15 & B& 18:53:04 & 23:11:42 & 4.3& 1		\\
2018 Dec 21 & B& 20:11:18 & 22:59:59 & 2.82& 0		\\
2018 Dec 22 & B& 12:20:33 & 23:22:55 & 11.03& 1		\\
2018 Dec 23 & B& 13:52:46 & 23:26:18 & 9.57& 4		\\
2018 Dec 27 & B& 16:35:30 & 22:16:12 & 5.68& 3		\\
2019 Jan 19 & B& 14:37:51 & 18:59:57 & 4.37& 2		\\
2019 Jan 20 & B& 13:35:37 & 18:59:48 & 5.4& 0		\\
2019 Jan 23 & B& 13:33:13 & 16:27:56 & 2.9& 0		\\
2019 Jan 27 & B& 14:32:42 & 18:59:55 & 4.45& 1		\\
2019 Jan 29 & B& 13:55:26 & 17:59:52 & 4.07& 0		\\
2019 Feb 11 & B& 12:59:35 & 16:54:58 & 3.92& 3		\\
2019 Feb 13 & B& 15:03:50 & 17:59:49 & 2.93& 1		\\
2019 Apr 03 & B& 14:00:01 & 15:58:52 & 1.97& 0		\\
2019 Apr 04 & B& 14:36:32 & 15:59:49 & 1.38& 0		\\
2019 Apr 09 & B& 14:23:17 & 15:59:44 & 1.6& 0		\\
2019 Apr 14 & B& 14:34:07 & 15:59:44 & 1.42& 1		\\
2019 Apr 15 & B& 14:33:53 & 15:59:53 & 1.43& 0		\\
  \noalign{\smallskip}\hline
\end{tabular}
\end{center}
\end{table}

\begin{table}[ht]
\begin{center}
\caption[]{ \centering {Log of spectrometry} }\label{Tab:t2}
\begin{tabular}{cccccc}
  \hline\hline\noalign{\smallskip}
Date & Start & End & Duration & Flare \\
(UT) & (UT) & (UT) & (Minute) & Number \\
  \hline\noalign{\smallskip}
2019 Feb 11  &  12:58:30 & 14:04:06 & 65 & 1 \\
  \noalign{\smallskip}\hline
\end{tabular}
\end{center}
\end{table}
%% Authors can give a citation as 'Michel et al. 1992'.
%% You may also use \cite, \citep and \citet for citation, and use Table~1 or Figure~1
%% and so forth. Using \ref and \label for cross-references of Tables/Figures
%% is a good way in adjusting/adding/removing text, tables or figures.

\section{Observations and data reduction} \label{sec:Observation}

\subsection{Photometric data} \label{subsec:Photometric}
The photometric observations of Ross 15 (RA 01:59:24, Dec +58:31:16) were executed by using the Ningbo Bureau of Education and Xinjiang Astronomical Observatory Telescope (NEXT). The NEXT is located at Nanshan station of Xinjiang Astronomical Observatory, China, with an aperture of 60 cm and a focal ratio of F/8. A FLI 230-42 CCD with 2048 $\times$ 2048 pixels is mounted on the prime focus of the telescope, providing a field of view of 22 $\times$ 22 arcmin$^{2}$. A set of Johnson-Cousins UBVRI filters is equipped for the broadband photometry. The limiting magnitude of the B band could reach to 15.4 mag in signal-to-noise ratio (SNR) = 20 for 10 seconds of exposure.

We carried out the photometric observations for Ross 15 on 26 nights, $\sim $128 hours in total, with the B filter from October, 2018 to April, 2019, using the NEXT. Table 1 shows the log of the photometry, including the observation date, the filter used, the start and end times of each observing section, the section duration, and the number of the detected flares in each section. The exposure time is 10 seconds and the readout time is $\sim $15 seconds, leading to a $\sim $25-second cadence for each observing section.

The photometric data is reduced by the software MaxIm DL 5.15 with the standard procedure of bias subtraction, dark subtraction and flat correction. The differential photometry is also performed with MaxIm DL 5.15. TYC 3696-453-1 (RA 01:59:05, Dec +58:29:09) is selected as a reference star. The formula, Error = 1.0857/SNR \citep{2016pglp.book.....W}, is used to compute the photometry error in magnitude, which is same with that of IRAF package.

\begin{figure}[ht]
\centering
\includegraphics[angle=0,width=110mm]{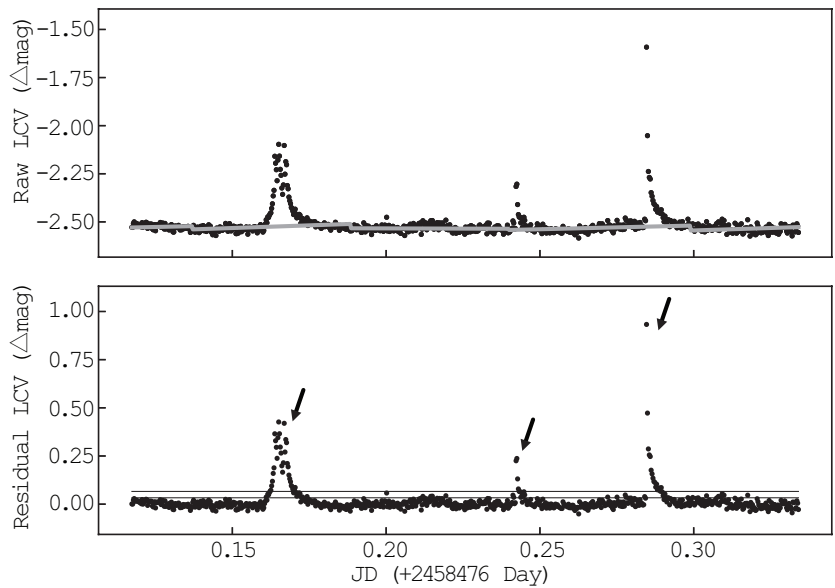}
\caption{The Raw LCV and the Residual LCV on 2018 December 23. The two light curves are marked with the black points. In the top panel, the thick gray line shows polynomial fit of the quiescent phase. The fitting shows some sudden changes as we fit the quiescent phase by dividing it into several parts in order to obtain a better fitting. In the bottom panel, the two black lines indicate the positions of 2 $\times$ SD (lower) and 4 $\times$ SD (higher), respectively. And the arrows show three identified flares. \label{fig:f1}}
\end{figure}
\subsection{Spectral data} \label{subsec:Spectral}
Simultaneously with the photometric observation on February 11, 2019, the high-cadence spectra of Ross 15 were obtained by using a 2.16-m telescope. This telescope is located at Xinglong station of the National Astronomical Observatories of China, which is mounted with the Beijing Faint Object Spectrograph and Camera (BFOSC). The wavelength coverage is from 3300 to 10000 \AA, and the resolution power is R = 1120 at 6500 \AA. The limiting magnitude of the telescope in the spectroscopic observations typically could reach V = 20 mag (SNR = 5) within 1 hour of exposure. More parameters of the telescope and the BFOSC can be found in \cite{2016PASP..128k5005F}.

Table 2 presents the log of the spectrometry. The integrated time is 10 seconds and the readout time is $\sim$5 seconds for each spectrum, leading to a $\sim$15-second cadence. A total of 255 spectra are obtained in 65 minutes of duration. BD+75D325 is selected as a standard star, and its spectrum is obtained with 60-second exposure time. A total of five flat images and one Fe-Ar lamp spectrum are obtained at the beginning of the whole observation.

All the spectra are reduced with IRAF packages\footnote{IRAF is distributed by the National Optical Astronomy Observatories (NOAO) in Tucson, Arizona, which is operated by the Association of Universities for Research in Astronomy, Inc., under cooperative agreement with the National Science Foundation.}. In order to achieve higher SNR for the spectra of Ross 15, every two spectra are combined. The combined spectra are then reduced in a standard procedure which involved zero subtraction, flat correction and spectrum extraction. Wavelength and flux calibration are performed with the Fe-Ar lamp spectrum and the spectrum of the standard star, respectively.

\begin{table}[h]
\begin{center}
\caption[]{ \centering {Parameters of flares} }\label{Tab:t33}
\begin{tabular}{cccccc}
  \hline\hline\noalign{\smallskip}
Flare &  T$_{rise}$      & T$_{decay}$ & T$_{total}$ & Amplitude & log E$_B$ \\
ID &  (Minute)    & (Minute) & (Minute) & (mag) &  \\
  \hline\noalign{\smallskip}
1	&	0.73 	&	3.70 	&	4.43 	&	0.16 	$\pm$	0.01 	&	30.12 	$\pm$	0.05 	\\
2	&	0.37 	&	2.20 	&	2.57 	&	0.06 	$\pm$	0.01 	&	29.45 	$\pm$	0.05 	\\
3	&	1.07 	&	8.55 	&	9.62 	&	0.26 	$\pm$	0.01 	&	30.37 	$\pm$	0.04 	\\
4	&	11.33 	&	15.35 	&	26.68 	&	0.27 	$\pm$	0.01 	&	30.40 	$\pm$	0.04 	\\
5	&	1.07 	&	6.27 	&	7.33 	&	0.20 	$\pm$	0.01 	&	29.75 	$\pm$	0.04 	\\
6	&	0.37 	&	11.22 	&	11.58 	&	0.35 	$\pm$	0.02 	&	30.49 	$\pm$	0.06 	\\
7	&	2.53 	&	9.87 	&	12.40 	&	0.06 	$\pm$	0.01 	&	30.15 	$\pm$	0.05 	\\
8	&	1.17 	&	6.23 	&	7.40 	&	0.38 	$\pm$	0.01 	&	30.70 	$\pm$	0.05 	\\
9	&	2.98 	&	6.68 	&	9.67 	&	0.28 	$\pm$	0.02 	&	30.70 	$\pm$	0.09 	\\
10	&	0.42 	&	2.97 	&	3.38 	&	0.08 	$\pm$	0.01 	&	30.00 	$\pm$	0.03 	\\
11	&	1.23 	&	2.88 	&	4.12 	&	0.32 	$\pm$	0.01 	&	30.16 	$\pm$	0.03 	\\
12	&	1.88 	&	17.23 	&	19.12 	&	1.21 	$\pm$	0.04 	&	31.30 	$\pm$	0.13 	\\
13	&	0.85 	&	1.75 	&	2.60 	&	0.16 	$\pm$	0.02 	&	29.93 	$\pm$	0.07 	\\
14	&	1.13 	&	1.10 	&	2.23 	&	0.16 	$\pm$	0.02 	&	30.15 	$\pm$	0.06 	\\
15	&	5.50 	&	11.08 	&	16.58 	&	0.43 	$\pm$	0.02 	&	30.98 	$\pm$	0.06 	\\
16	&	0.73 	&	3.68 	&	4.42 	&	0.24 	$\pm$	0.02 	&	30.10 	$\pm$	0.06 	\\
17	&	0.45 	&	9.95 	&	10.40 	&	0.93 	$\pm$	0.02 	&	30.77 	$\pm$	0.06 	\\
18	&	1.22 	&	14.63 	&	15.85 	&	0.96 	$\pm$	0.02 	&	31.09 	$\pm$	0.06 	\\
19	&	0.75 	&	1.50 	&	2.25 	&	0.13 	$\pm$	0.02 	&	30.05 	$\pm$	0.06 	\\
20	&	0.73 	&	6.07 	&	6.80 	&	0.12 	$\pm$	0.02 	&	30.38 	$\pm$	0.06 	\\
21	&	0.87 	&	1.33 	&	2.20 	&	0.12 	$\pm$	0.01 	&	29.76 	$\pm$	0.05 	\\
22	&	0.87 	&	7.85 	&	8.72 	&	1.04 	$\pm$	0.01 	&	30.93 	$\pm$	0.05 	\\
23	&	1.28 	&	8.58 	&	9.87 	&	0.92 	$\pm$	0.02 	&	31.02 	$\pm$	0.08 	\\
24	&	4.12 	&	28.42 	&	32.53 	&	1.07 	$\pm$	0.01 	&	31.53 	$\pm$	0.03 	\\
25	&	0.83 	&	10.68 	&	11.52 	&	0.17 	$\pm$	0.01 	&	30.50 	$\pm$	0.03 	\\
26	&	0.80 	&	5.35 	&	6.15 	&	0.17 	$\pm$	0.01 	&	30.26 	$\pm$	0.03 	\\
27	&	0.43 	&	1.68 	&	2.12 	&	0.39 	$\pm$	0.01 	&	30.09 	$\pm$	0.04 	\\
28	&	0.42 	&	6.03 	&	6.45 	&	0.23 	$\pm$	0.02 	&	30.19 	$\pm$	0.06 	\\

  \noalign{\smallskip}\hline
\end{tabular}
\end{center}
\end{table}

\section{Data Analysis}\label{sec:Analysis}
\subsection{Photometric analysis} \label{subsec:PhotometricAnalysis}
A program has been written to detect the flares from the light curve of Ross 15 on each night in an automatic manner. Through the program, in the first step, the light curve from a night (Raw LCV) is fitted iteratively by smoothing average to exclude outliers in order to obtain a quiescent phase \citep{2014ApJ...797..121H,2016ApJ...829L..31D,2017ApJ...849...36Y,2019ApJS..241...29Y}. In the second step, the quiescent phase is fitted by a polynomial fitting, and then the Raw LCV is subtracted by the polynomial fitting to get a subtracted light curve (Residual LCV). In the final step, the Residual LCV is checked to get flare candidates with the similar criteria applied in previous research \citep{2011ASPC..448..197H,2012PASP..124..545H,2014ApJ...797..121H} (see Figure 1). The criteria are as follows: (1) In a light curve of a flare, there are at least seven consecutive measurements; (2) These measurements are more than two times of the standard deviation (SD) of the quiescent phase and at least one of them is more than four times of the SD. Beside the two criteria, the profile of the light curve of each flare candidate is also checked through a graphic interface to confirm that the profile consists of an impulsive rise (relatively short) and an exponential decay (relatively long).

\begin{figure}[h]
\centering
\includegraphics[angle=0,width=75mm]{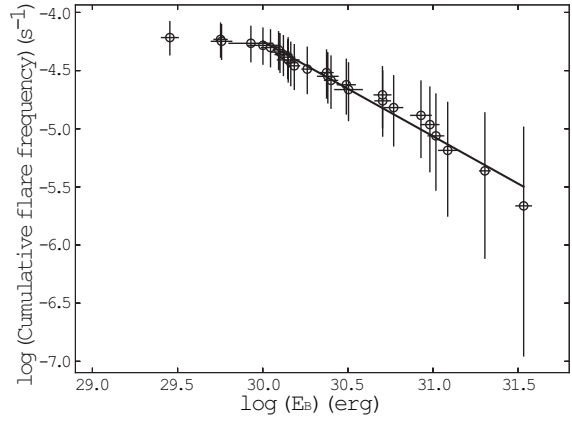}
\caption{Cumulative flare frequency distribution versus flare energy for the 28 events observed on Ross 15 in B bandpass. The open black circles represent the cumulative flare frequency at each flare energy. The vertical error bars show the confidence interval of 95\% and the horizontal indicate the error of energy. The black line indicates the least-squares power-law fit, log(Flare frequency) = - 0.82$^{+0.16}_{-0.16}$ log(E$_B$) + 20.23$^{+4.05}_{-4.05}$. \label{fig:f2}}
\end{figure}

The cumulative flare frequency distribution (FFD) is a diagram of cumulative flare frequency (log number of flares per hour with energy greater than that of a flare) versus flare energy \citep{1972Ap&SS..19...75G,1976ApJS...30...85L,2018ApJ...858...55P}. Figure 2 shows the calculated FFD for the B-band flares on Ross 15 in this work. Following the method applied in \cite{2014ApJ...797..121H}, B-band flare energy is computed with the equivalent duration \citep{1972Ap&SS..19...75G} of a flare multiplied by the quiescent luminosity of the star. The equivalent duration is defined as the amount of the time that the star would take in its quiescent state to release the same amount of energy released during a flare, and is calculated as the time integral of F$_{f}$(t)/F$_{0}$, where F$_{f}$(t) is the flux of the flare and F$_{0}$ is the flux of the star in the quiescent state \citep[see][]{1972Ap&SS..19...75G,2012PASP..124..545H,2014ApJ...797..121H}. The B-band flux is estimated by convolving the transmission of B-band filter with the quiescent spectrum. Then, the quiescent luminosity in B band is calculated with 4$\pi$d$^2$ multiplied by the B-band flux of the quiescent state of Ross 15, where d = 13.1348 $\pm$ 0.0093 pc is the distance from the star to the earth \citep{2018yCat..36160012G}. We estimate the quiescent luminosity of the star as 1.26 $\times$ 10$^{29}$ erg s$^{-1}$. The confidence interval of 95\% for flare frequency is calculated using the Poisson confidence intervals \citep{1986ApJ...303..336G, 2016ApJ...829L..31D}. The errors for the flare energies are computed with the errors of the photometry. For all the flares detected, we also calculated the amplitudes, the times of rise and decay, and the total duration of the flares (see Table 3).

\begin{figure}[h]
\centering
\includegraphics[width=120mm, angle=0]{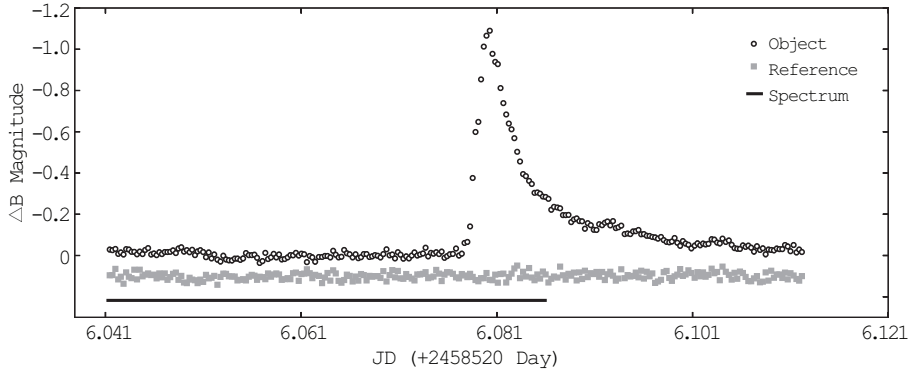}
\caption{ The B-band light curve of the 1.07-mag flare and the duration of the spectroscopic observations. The light curve of Ross 15 is marked with open circles. Gray solid squares indicate the light curve of the reference star. The 1$\sigma$ errors of the photometry are less than 0.01 mag. The error bars are smaller than the sizes of the symbols in both curves. The duration of the spectrometry is marked with thick black line. }
\label{Fig:f6}
\end{figure}

\begin{figure}[h]
\centering
\includegraphics[angle=0,width=70mm]{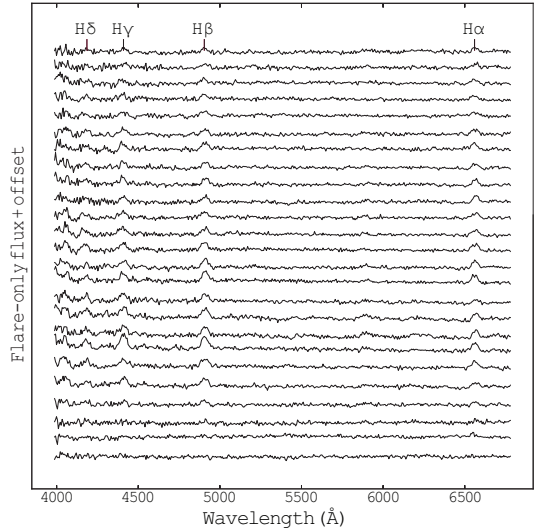}
\caption{The flare-only spectra during the 1.07-mag flare. To be clear, the spectra are shifted vertically and plotted from bottom to top by the observational order. H$\alpha$, H$\beta$, H$\gamma$ and H$\delta$ emission lines are indicated on the top.\label{fig:f3}}
\end{figure}

\subsection{Spectroscopic analysis} \label{subsec:SpectroscopicAnalysis}

One flare is detected during the simultaneous spectral and photometric observations. The photometric amplitude of the flare is up to 1.07 mag, so we note this flare as 1.07-mag flare hereafter (see Figure 3). In this work, we define 'pre-flare' as the quiescent phase just before a flare. Intensities of emission lines and continuum are calculated to track their evolutions during the pre-flare and flare phase. All the combined spectra in the pre-flare phase are averaged to build a template. The spectra in the pre-flare and flare phase are subtracted by the template, deriving flare-only spectra \citep{2010ApJ...714L..98K}, which present H$\alpha$ (6563 \AA), H$\beta$ (4861 \AA), H$\gamma$ (4341 \AA) and H$\delta$ (4102 \AA) as shown in Figure 4. Wide wavelength ranges, 6537 -- 6583 {\AA}, 4887 -- 4933 {\AA}, 4385 -- 4431 {\AA} and 4149 -- 4195 {\AA} are applied to calculate the intensities of H$\alpha$, H$\beta$, H$\gamma$ and H$\delta$, respectively. The range of 4218 - 4328 {\AA} is used to obtain the intensity of the continuum. The accumulated flux in the wavelength range of the continuum or each of the emission lines in a flare-only spectrum is divided by the wavelength to get the intensity. All intensities calculated are presented in Figure 5.

\begin{figure}[h]
\centering
\includegraphics[angle=0,width=70mm]{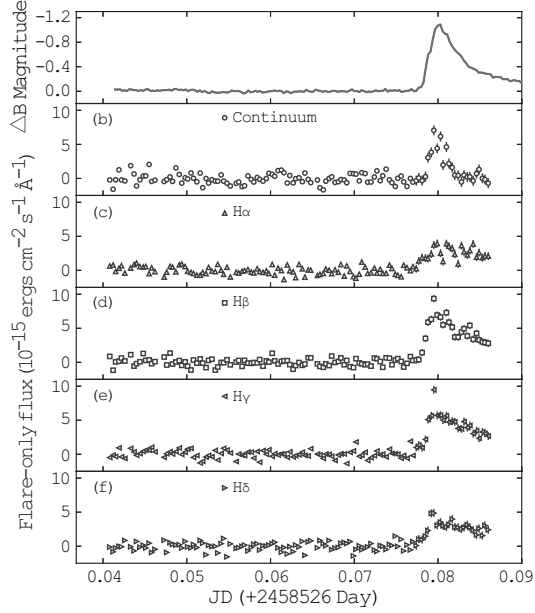}
\caption{Panel a: The light curve of the 1.07-mag flare. Panel b to f: The intensity evolutions of the continuum, H$\alpha$, H$\beta$, H$\gamma$ and H$\delta$. Error bars on all panels represent the root mean square of the intensities in the pre-flare phase. \label{fig:f4}}
\end{figure}

\begin{figure}[h]
\centering
\includegraphics[angle=0,width=70mm]{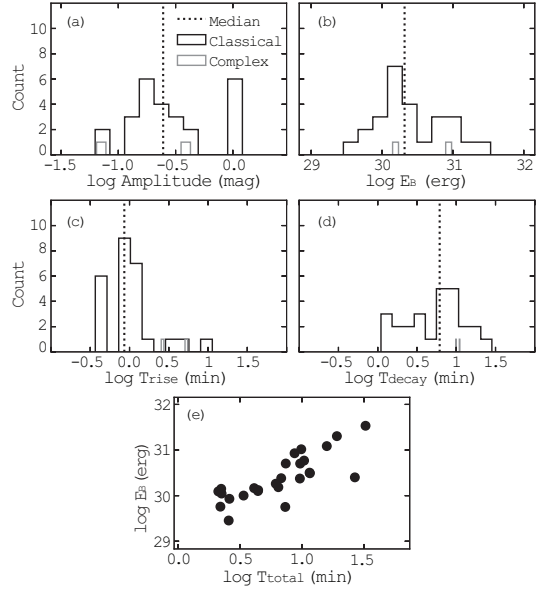}
\caption{Panel a to d: Histograms of flare parameters. In all the panels, the black and gray lines indicate the classical and complex flares, respectively, and the vertical dotted lines show the medians of the parameters of the classical flares. Panel e: Distribution of the flare energies in B band versus total durations.\label{fig:f5}}
\end{figure}

\section{Results}\label{sec:Results}

Through the method described in Section 3.1, a total of 28 B-band flares are identified in the $\sim$128 hours of photometric observations, leading to a flare rate of 0.22$^{+0.04}_{-0.04}$ hour$^{-1}$ through our observations. Among them, 26 flares are classical (one peak) and 2 are complex (two or more peaks). Table 3 gives the parameters of each flare, including the flare ID, the times of the rise and decay, the total duration of a flare, the peak amplitude in magnitude and the B-band energy in log. In Figure 6, the panel a to d give the histograms of the amplitudes, the energies, the times of the rise and decay. The median values of the four parameters are 0.25 mag, 10$^{30.5}$ erg, 0.25 minutes and 6.2 minutes respectively.

Following the approach of \cite{2016ApJ...829..129S}, we show in the panel e of Figure 6 the distribution of the flare energies versus total durations. The correlation coefficient between the two parameters is 0.8. The B-band energy range of our detected flares is from 10$^{29.5}$ to 10$^{31.5}$ erg. The FFD of Ross 15 is shown in Figure 2, and is fitted by the least-squares power law of log(Flare frequency) = - 0.82$^{+0.16}_{-0.16}$ log(E$_B$) + 20.23$^{+4.05}_{-4.05}$.

Figure 3 shows the light curve of the 1.07-mag flare and the observing duration of the spectra of Ross 15. This flare is the most energetic one (10$^{31.5}$ erg) in our sample. In the observing duration, a total of 204 spectra are obtained in pre-flare phase and 51 during the flare phase, resulting in 127 combined spectra in total. In pre-flare phase, the minimum SNR of the combined spectra at H$\alpha$, H$\beta$, H$\gamma$ and H$\delta$ are 38, 23, 12 and 14, respectively. The intensity evolutions of continuum and the emission lines are presented in Figure 5, which shows significant enhancements of them during the flare. Figure 7 illustrates the distributions of the intensities of the continuum versus that of H$\alpha$, H$\beta$, H$\gamma$ and H$\delta$ during the flare phase, with the correlation coefficients of 0.6, 0.8, 0.7 and 0.5, respectively.

\begin{figure}[h]
\centering
\includegraphics[angle=0,width=70mm]{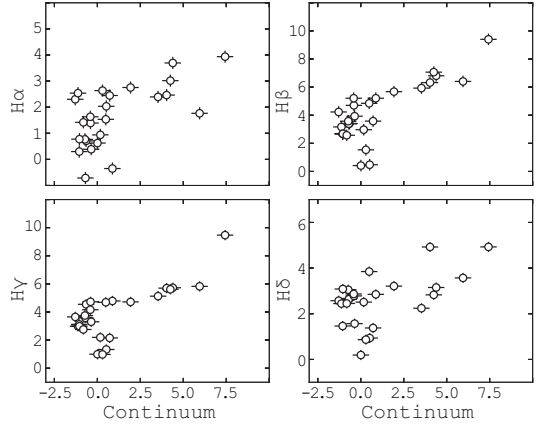}
\caption{Intensity distributions between the continuum and Balmer lines during the flare phase. To be clear, all the intensity values are divided by 10$^{-15}$ ergs cm$^{-2}$ s$^{-1}$ \AA$^{-1}$, and marked with open circles. The vertical error bars in each panel represent the standard error of the intensities of each emission line, and the horizontal represent that of the continuum. \label{fig:f7}}
\end{figure}

\begin{figure}[h]
\centering
\includegraphics[angle=0,width=70mm]{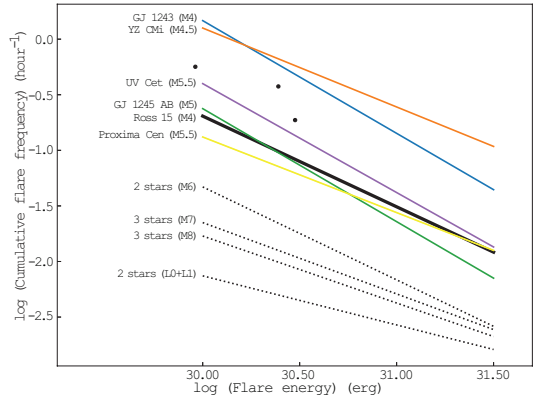}
\caption{Comparison of flare rates of Ross 15 (thick black line) with that of GJ 1243 (M4, blue line), YZ CMi (M4.5, red line), UV Cet (M5.5, purple line), GJ 1245 AB (M5, green line), Proxima Cen (M5.5, yellow line) and average flare rates of different spectral types (M6 to L0+L1, dotted lines). The FFDs of YZ CMi and UV Cet are taken from \cite{1976ApJS...30...85L}, and that of GJ 1243, GJ 1245 AB and Proxima Cen are taken from \cite{2014ApJ...797..121H}, \cite{2015ApJ...800...95L} and \cite{2016ApJ...829L..31D}, respectively. The average flare rates of M6 to L0+L1 are adopted from \cite{2018ApJ...858...55P}. Star name/numbers of stars mixed and spectral type are indicated on the left of each line. The FFD of Ross 15 in \cite{1977A&AS...30..113P} is also plotted with black points.} \label{fig:f8}
\end{figure}

\section{Discussion}\label{sec:Discussion}

On the study of flares of Ross 15, \cite{1977A&AS...30..113P} is the only published paper so far, in which a flare rate of 0.51$^{+0.29}_{-0.29}$ hour$^{-1}$ is given based on three B-band flares detected in 5.85 hours of photometric observations using a 60-cm telescope. In our work, with a telescope the same size, a more accurate B-band flare rate of 0.22$^{+0.04}_{-0.04}$ hour$^{-1}$ is detected in 128 hours of observations. \cite{1977A&AS...30..113P} gives a flare rate about twice that of ours, most likely due to the relatively small number of flare samples observed in a relatively short observation time, resulting in a large statistical deviation of flare rate in that work.

The relatively large number of flare samples in our work allows us to analyze the flare energy in B band, especially to obtain the FFD of Ross 15. In Figure 8, within the energy range of the flares we detected, we compare the FFD of this star with that of different spectral types. M4 type includes GJ 1243 (M4) and YZ CMi (M4.5), and M5 includes UV Cet (M5.5), GJ 1245 AB (M5) and Proxima Cen (M5.5). For M6 to L0+L1, the average FFDs are computed by mixing that of targets of similar spectral types \citep{2018ApJ...858...55P}. The FFDs of YZ CMi and UV Cet are reported using U-band energies of flares \citep{1976ApJS...30...85L}, and those of GJ 1243 and GJ 1245 AB applying Kepler energies \citep{2014ApJ...797..121H,2015ApJ...800...95L}. The FFD of Proxima Cen is reported adopting MOST\footnote{The Microvariablity and Oscillations of Stars mission, whose details can be found in \cite{2003PASP..115.1023W}.} energies of flares \citep{2016ApJ...829L..31D}. The average flare rates of spectral types $\geq$ M6 are taken from \cite{2018ApJ...858...55P}, which are also computed from Kepler energies of flares. As shown in Figure 8, the flare rates of Ross 15 are lower than that of YZ CMi and GJ 1243, and approximately in the middle of that of the three M5-type stars, and higher than those of spectral types $\geq$ M6. For M-type stars, the statistical results of the rotation - activity relation from several studies indicate that the stars with the shorter rotation periods tend to have higher flare activity \citep[see][]{2019ApJ...873...97L,2019ApJS..241...29Y}. The M4-type stars, YZ CMi and GJ 1243, have the rapid rotation periods, 2.77729 and 0.5927 $\pm$ 0.0002 day respectively \citep{2014ApJ...797..121H,2017AstBu..72..178B}. Their rotation periods are probably shorter than that of Ross 15, resulting in the higher flare rates in the FFD of them than that of our object. Figure 8 also shows that the FFD slope of Ross 15 is comparable with that of M4 to M6 stars, and steeper than that of spectral types $\geq$ M7.

Simultaneous high-cadence spectroscopic and photometric observations reveal the properties of intensity evolutions of the continuum and emission lines during the 1.07-mag flare of Ross 15. The intensities of Balmer lines (H$\alpha$, H$\beta$, H$\gamma$ and H$\delta$) are enhanced significantly during the flare. Optical wavelength observations in previous researches also show that the intensities of H$\alpha$ and H$\beta$ are enhanced during the flares of EV Lac \citep{2006Ap.....49..488M,2018PASJ...70...62H}. The enhancements of the Balmer lines indicate that the chromospheric activities of the flare stars occur in the middle chromosphere \citep{2000A&AS..146..103M,2011ASPC..451..123Z,2015RAA....15..252Z}. In addition to the Balmer lines, several studies have shown that Ca {\tiny II} H \& K, Na {\tiny I} D and He {\tiny I} lines also exhibit enhanced radiation during flares \citep{2013A&A...560A..69L,2013ApJS..207...15K,2016ApJ...820...95K}. Near- and Far-ultraviolet wavelength observations of the Hubble Space Telescope show that the emission lines and continuum also display enhancements during the flares of GJ 1243 and GJ 674 \citep{2019ApJ...871..167K,2019ApJ...871L..26F}.

\section{Summary}\label{sec:Summary}
Photometric and spectroscopic observations are conducted for Ross 15. A total of 28 B-band flares are detected in 128-hour photometric observations, bring about a flare rate of 0.22$^{+0.04}_{-0.04}$ hour$^{-1}$, more accurate than that provided in previous work. The energy range of the B-band flare (10$^{29.5}$ - 10$^{31.5}$ erg) and the FFD of Ross 15 are given for the first time. We compare the FFD with that of different spectral types in the same energy range. The comparison shows that the FFD of Ross 15 are lower than that of GJ 1243 and YZ CMi, approximately in the middle of that of the three M5-type stars and higher than the average FFDs of spectral types $\geq$ M6. Simultaneous high-cadence spectroscopic and photometric observations are performed for the first time to Ross 15, leading to detection of the most energetic flare in our sample. The intensity evolutions of the continuum and Balmer lines are tracked for the flare. Significant enhancements of intensities of the continuum and Balmer lines are shown during the flare, and the correlation analysis indicates that there are significant correlations between them, which is same with that of the other deeply studied flare stars of the similar spectral type.

\begin{acknowledgements}
This research is supported by the National Natural Science Foundation of China under grant No. 11873081, and also partially supported by the Open Project Program of the Key Laboratory of Optical Astronomy, National Astronomical Observatories, CAS. We acknowledge the support of the staff of the Xinglong 2.16-m telescope.
\end{acknowledgements}

\bibliographystyle{raa}
\bibliography{ms2020-0369}

\end{document}